\documentclass[final,5p,times,twocolumn,sort&compress,nopreprintline]{elsarticle}
\biboptions{square,sort&compress,comma,numbers}

\usepackage[colorlinks]{hyperref}
\usepackage[table,xcdraw,dvipsnames]{xcolor}
\usepackage[normalem]{ulem}

\usepackage[latin1]{inputenc}
\usepackage[english]{babel}
\usepackage{amssymb,amsmath,amsthm,cancel,graphicx,xcolor}
\usepackage{booktabs}
\usepackage{mathrsfs}

\usepackage{color}

\let\originalleft\left
\let\originalright\right
\renewcommand{\left}{\mathopen{}\mathclose\bgroup\originalleft}
\renewcommand{\right}{\aftergroup\egroup\originalright}

\newcommand{\mX}{\mathcal{X}}

\def\beq{\begin{equation}}  
\def\eeq{\end{equation}}
\def\({\left(}
\def\){\right)}
\def\[{\left[}
\def\]{\right]}

\def\eq#1{{Eq.~(\ref{#1})}}
\def\eqs#1#2{{Eqs.~(\ref{#1})-(\ref{#2})}}

\def\Table#1{{Table~\ref{#1}}}

\def\sect#1{{Sect.~\ref{#1}}}

\def\Tr{\mbox{Tr}\,}
\def\tr{\mbox{Tr}\,}

\def\wt#1{\widetilde{#1}}

\renewcommand{\bar}{\overline}

\newcommand{\vckm}{V_{\text{CKM}}}

\newcommand{\X}{{\cal X}}
\newcommand{\U}{{\rm U}}

\newcommand{\qL}{{q_L}}
\newcommand{\uR}{{u_R}}
\newcommand{\dR}{{d_R}}
\newcommand{\lL}{{\ell_L}}
\newcommand{\eR}{{e_R}}

\newcommand{\GeV}{\,\mathrm{GeV}}

 \renewcommand{\b}{\beta}

\newcommand{\SU}{{\rm SU}}

\newcommand{\BSM}{{\rm BSM}}

\newcommand{\1}{{\textbf{1}}}

\journal{arXiv}

\begin{document}

\begin{frontmatter}



\title{
Renormalization group effects in astrophobic axion models
}


\author[label1,label2]{Luca Di Luzio} 
\author[label3]{Federico Mescia} 
\author[label4]{Enrico Nardi} 
\author[label3]{Shohei Okawa} 
\address[label1]{Dipartimento di Fisica e Astronomia `G.~Galilei', Universit\`a di Padova, Italy}
\address[label2]{Istituto Nazionale di Fisica Nucleare, Sezione di Padova, Padova, Italy}
\address[label3]{Departament de F\'isica Qu\`antica i Astrof\'isica, Institut de Ci\`encies del Cosmos (ICCUB), \\ 
Universitat de Barcelona, Mart\'i i Franqu\`es 1, E-08028 Barcelona, Spain}
\address[label4]{Istituto Nazionale di Fisica Nucleare, Laboratori Nazionali di Frascati, C.P.~13, 00044 Frascati, Italy}

\begin{abstract}
It has been recently pointed out that in certain axion models it is possible to suppress simultaneously both the axion couplings to nucleons and  electrons, realising the so-called {\it astrophobic axion} scenarios, wherein the tight bounds from SN1987A and from  stellar evolution of red giants and white dwarfs are greatly relaxed.
So far, however, the conditions for realising astrophobia have only been set out in tree-level analyses. 
Here we study whether these conditions can still be consistently implemented once renormalization group 
effects are included in the running of axion couplings. 
We find that axion astrophobia keeps holding, albeit within fairly different 
parameter space regions, and we provide 
analytical insights into this result.
Given that astrophobic axion models generally feature flavour violating axion couplings, 
we also assess the impact of renormalization group effects on axion-mediated flavour violating observables.
\end{abstract}





\end{frontmatter}

\thispagestyle{plain}


\section{Introduction}
\label{sec:intro}
Non-universal axion model, in which 
the Peccei-Quinn (PQ) symmetry 
$\U(1)_{\rm PQ}$   
\cite{Peccei:1977hh,Peccei:1977ur,Weinberg:1977ma,Wilczek:1977pj}
acts on the different Standard Model (SM) fermions in a generation-dependent way, 
have been often considered in frameworks addressing the 
SM flavour puzzle (see e.g.~Refs.~\cite{Ema:2016ops,Calibbi:2016hwq,Bjorkeroth:2018ipq}), 
as well as in more phenomenological 
contexts. For instance, it was recently pointed out in Ref.~\cite{DiLuzio:2017ogq} that 
in variants of Dine-Fischler-Srednicki-Zhitnitsky (DFSZ)
\cite{Dine:1981rt,Zhitnitsky:1980tq} models with two Higgs doublets (2HDM)
the non-universality of  the SM quarks PQ charges is a necessary ingredient to allow a 
simultaneous suppression of the axion coupling both to protons and neutrons.
Nucleophobia can then be obtained in parameter space regions in which the ratio
of the two Higgs vacuum expectation values (VEVs) satisfies certain conditions. 
This allows to relax the tight astrophysical  bounds on the decay constant $f_a$
(or on the axion mass $m_a$) 
 from  Supernova (SN) 1987A. 
Still, the bounds are only marginally loosened  
because in DFSZ-like  models axion couplings to electrons are generically of $\mathcal{O}(1/f_a)$, and then limits from white dwarfs and red giants stars evolution, which are only moderately weaker than the SN1987A bound 
(see e.g.~Ref.~\cite{DiLuzio:2021ysg} for 
a recent review) still apply. 
Axion-electron decoupling can be either 
obtained at the price of an extra tuning 
with the flavour structure of the lepton 
rotation matrices \cite{DiLuzio:2017ogq} or, 
more elegantly, it can be  
implemented together with nucleophobia, and without extra tuning, in a three Higgs 
doublets model (3HDM), as detailed in Ref.~\cite{Bjorkeroth:2019jtx}.  
In Refs.~\cite{DiLuzio:2017ogq,Bjorkeroth:2019jtx} the conditions for nucleo/electrophobia were formulated in terms of tree-level relations 
(up to small QCD running 
effects \cite{diCortona:2015ldu}) and  it is then mandatory to question whether the resulting suppression of the axion couplings to nucleons and electron can survive  after  including the effects of radiative corrections.   

The full one-loop anomalous dimensions for the $d=5$ axion effective 
Lagrangian have been recently computed in Refs.~\cite{Choi:2017gpf,Bauer:2020jbp}, 
while running effects have been systematically investigated, within canonical axion models, in Ref.~\cite{Choi:2021kuy}. 
For related efforts to include 
loop effects on flavour-violating axion couplings, 
with a non-trivial dependence from the UV completion, 
see Ref.~\cite{Alonso-Alvarez:2021ett}.
The purpose of this work is to extend the analysis of the 
running axion couplings  to non-universal axion models, and to
assess, in particular, the radiative stability 
under the renormalization group (RG) evolution of the nucleo/electrophobic conditions 
set out in Refs.~\cite{DiLuzio:2017ogq,Bjorkeroth:2019jtx}. 
A remarkable consequence of non-universal axion models is the generic occurence  
of flavour-violating axion couplings, which can be tested in low-energy flavour-changing 
process, such as e.g.~$K \to \pi a$,  
that will be probed at current and future experimental facilities 
\cite{NA62:2017rwk,KOTO:2018dsc,KLEVERProject:2019aks}. We hence complement our  study 
by assessing the relevance of  running effects for flavour off-diagonal axion couplings.

\section{Astrophobic axions}
\label{sec:nonUnivAxion}

We  focus first on a specific non-universal axion model introduced in 
Ref.~\cite{Bjorkeroth:2019jtx}, wherein the nucleo and electrophobic conditions
can be elegantly realised within certain regions of the parameter space spanned by the ratios between the VEVs of the Higgs doublets that couple to SM fermions.

The model features three Higgs doublets $H_{1,2,3}$ (hence we will label 
it as 3HDM) and a SM singlet complex scalar $\Phi$. 
Under the SM gauge group $\SU(3)_C\times \SU(2)_L\times \U(1)_Y$ the 
quantum numbers  of the scalars are $H_{1,2,3} \sim (1,2,-1/2)$ and   
$\Phi \sim (1,1,0)$.
The SM quarks couple to the first two doublets $H_{1,2}$ and their PQ charges  
are characterized by a 2+1 structure, namely the first two generations replicate 
the same set of charges, while 
the PQ charges of the third generation differ.
The $\U(1)_{\rm PQ}$ charges are chosen  in such a way that all the entries in the up- 
and down-type quark Yukawa matrices are allowed, so that there are no texture zeros.
In contrast,  all  the leptons  couple to the third doublet $H_3$ and feature universal 
PQ charges.\footnote{An alternative 
Higgs configuration 
in the lepton sector, leading to a moderately 
photophobic axion, is discussed in Ref.~\cite{Lucente:2022vuo}.}
The Yukawa sector of the  model contains the following operators: 
\begin{align}
&\bar q_1  u_1 H_1\, ,\ \:\bar q_3  u_3 H_{2}\, , \ \:\bar q_1  u_3 H_{1}\, , \ \:\bar q_3  u_1 H_{2}\, , \nonumber  \\
  \label{eq:m1}
&\bar q_1  d_1 \tilde H_2\, , \ \:\bar q_3  d_3 \tilde H_{1}\, ,\ \:\bar q_1  d_3 \tilde H_{2}\, ,\ \:\bar q_3  d_1 \tilde H_{1}\, , \nonumber \\
&\bar \ell_i  e_j \tilde H_3\, , \quad i,\,j = 1,2,3\,,
\end{align}
where 
$\tilde H_{1,2,3} = i \sigma_2 H^*_{1,2,3}$.
Note that the generation label ``1'' for quarks denotes both the first and second generation,  which  by assumption have the same PQ charges.

We are interested in the axion couplings to the proton, 
neutron and electron, which are defined via the effective interaction
\begin{equation}
\label{eq:axionN}
    \frac{C_\psi}{2f_a} \partial_\mu a \, \bar{\psi} \gamma^\mu \gamma_5 \psi ,
\end{equation}
with $\psi=p,n,e$, $f_a = f/(2 N)$ where $f_a$ is the 
axion decay constant, $f$ is the scale at which the PQ symmetry is broken, and $2 N$ is the coefficient of the PQ-QCD anomaly.\footnote{For uniformity of notation  with  studies of running  axion  couplings~\cite{Choi:2017gpf,Bauer:2020jbp,Choi:2021kuy} 
in \ref{sec:AxionEFT} we will denote the  anomaly coefficient as $c_G = 2 N$.}
The fundamental couplings $C_q$ of the axion
to the quarks $q=u,d,\dots$ are also defined by 
\eq{eq:axionN} with the replacement 
$\psi \to q$. 
$C_{p,n}$ can be expressed in terms of $ C_q $ using non-perturbative inputs from nucleon matrix
elements (see e.g.~\cite{diCortona:2015ldu}). 
For later purposes 
it is more convenient to consider the  two linear combinations:
\begin{align} 
\label{eq:CppCn}
    C_p +C_n &=  
    0.52 \,\left( C_u  + C_d  - 1\right) - 2 \delta_s ,
\\
\label{eq:CpmCn}
    C_p -C_n &= 
    1.28\, \left(C_u  - C_d  - f_{ud}\right) , 
\end{align}
where the right hand sides are obtained by using the expressions for 
$C_{p,n}$ given in \eqs{eq:CpAppB}{eq:CnAppB}. 
In \eq{eq:CpmCn}  $f_{ud} = f_u - f_d$, where 
$ f_{u,d} = m_{d,u}/(m_d+m_u)$ are the model-independent contributions 
induced by the axion coupling to gluons 
in the physical basis in which the axion is not mixed with $\pi^0$. 
In \eq{eq:CppCn} 
is a small $\mathcal{O}(5\%)$ correction dominated by the $s$-quark contribution (see \ref{sec:AxionEFT}).
Neglecting $\delta_s$, the approximate conditions 
for astrophobia are: 
\begin{align} 
\label{eq:nucleo1}
C_u  + C_d  &\approx 1 \, ,  \\
\label{eq:nucleo2}
C_u  - C_d  &\approx f_{ud} \approx \frac{1}{3} \, ,  \\
\label{eq:electro}
C_e &\approx0\,.  
\end{align}
At the tree level, the relevant couplings $C^0_{u,d} = (\mX_{u_1,d_1}-\mX_{q_1})/(2N)$  and $C^0_{e}=(\mX_{e}-\mX_{\ell})/(2N)$
can be read off from the Yukawa operators in \eq{eq:m1}.
In terms of the PQ charges $\mX_{1,2,3}$
of the three Higgs doublets they read \cite{DiLuzio:2020wdo}
\begin{equation}
\label{eq:C0}
    C^0_u = -\frac{\mX_1}{2N}\,,\quad    C^0_d = \frac{\mX_2}{2N}\,, \quad
       C^0_t = -\frac{\mX_2}{2N}\,,\quad           C^0_e = \frac{\mX_3}{2N}\,,
\end{equation}
where for later convenience we have listed also the top-quark coupling $  C^0_t$.\footnote{
In \eq{eq:C0} we have neglected  possible corrections to the diagonal quark couplings arising 
from fermion mixing. 
Throughout this paper we will assume that 
these mixing corrections are negligible.}
Due to the particular 2+1 structure 
of the quarks  PQ charges, the  contribution to the PQ anomaly of the third generation cancels against
the contribution of one of the two light 
generations, and it is then straightforward to obtain  $2N=\sum_i(\mX_{u_i}+\mX_{d_i} - 2 \mX_{q_i})=\mX_2-\mX_1$.
This implies that, at tree level, the first condition 
for nucleophobia \eq{eq:nucleo1}  is always 
satisfied. 

Consider now the following terms in the scalar potential, which  are needed to break the $\U(1)^4$ rephasing symmetry of the  kinetic terms of the four scalars  down to $\U(1)_{\rm PQ}\times \U(1)_Y$:\,\footnote{Different choices for the scalar operators are  possible, but they do not allow to satisfy simultaneously the  nucleo and electrophobic conditions (see Ref.~\cite{Bjorkeroth:2019jtx}).}
\begin{align}
\label{eq:scalars}
    H^\dagger_3 H_1 \Phi^2 +  H^\dagger_3 H_2 \Phi^\dagger\,.
\end{align}
Normalizing the charges to $\mX_\Phi=1$ we derive the conditions: 
\begin{align}
 \mX_1 = \mX_3 -2\,,  \qquad   
    \label{eq:4conditions4}
 \mX_2 = \mX_3 +1\,, 
\end{align}
which yield $2N=\mX_2-\mX_1=3$.
Substituting the values of $\mX_{1,2}$ 
in \eqs{eq:nucleo2}{eq:electro}
we obtain that, in terms of tree-level couplings, astrophobia can be realised if the following conditions on $\mX_3$ can be simultaneously satisfied:
\begin{align}
\label{eq:X3}
    \mX_3  = 
    \frac{1}{2} (1-3 f_{ud}), \qquad
            \mX_3  = 0\,.
\end{align}
It is a fortunate numerical
accident that the actual value of $f_{ud}$ is indeed very close to $1/3$ (corresponding to $m_d/m_u\approx 2$) so that nucleophobia and electrophobia are mutually compatible.

As a final step let us consider the PQ-hypercharge orthogonality condition. Let us 
 parametrise the VEVs as
$v_1=v c_1 c_2,\ v_2 = v s_1 c_2 ,\ v_3 = v s_2$
with $v^2 = v_1^2 + v_2^2 + v^2_3 \simeq (246 \, \text{GeV})^2$,
$c_1 \equiv\cos\beta_1$, $c_2 \equiv \cos\beta_2$,
etc. 
By using \eq{eq:4conditions4} we obtain
\begin{equation}
\label{eq:PQ-Y}
\sum_{i=1,2,3} \mX_i v^2_i = 0 \quad \Rightarrow \quad 
\mX_3 = (3 c_1^2 - 1)c_2^2 \,.
\end{equation}
The condition $\mX_3\approx 0$ 
then selects a certain region in the 
$(\beta_1,\beta_2)$ plane where the tree level
axion couplings to 
nucleons and electrons  
can be conveniently suppressed (see Fig.\,1 in Ref.~\cite{Bjorkeroth:2019jtx}).

A simpler astrophobic model  with only two Higgs doublets $H_{1,2}$
in which the  2+1 structure is extended also to the leptons 
was originally presented in   Ref.~\cite{DiLuzio:2017ogq} 
(see also  Ref.~\cite{DiLuzio:2021ysg}) and it was labeled ``model M1".
The Yukawa terms for the quarks are as in~\eq{eq:m1}, while the lepton Yukawas, the operators involving the two scalar doublets and the singlet $\Phi$, and the
PQ-hypercharge orthogonality condition  now involving only two Higgs doublets 
(i.e.~$\beta_2=0$) read, respectively:
\begin{align}
\label{eq:2HDMleptons}
 &\bar \ell_1  e_1 \tilde H_{1}\, , \ \   \bar \ell_3  e_3 \tilde H_{2}\, , \ \    \bar \ell_1   e_3 \tilde H_{1}\, ,\ \  \bar \ell_3  e_1 \tilde H_{2}\, , \\
& H_2^\dagger  H_1 \Phi   \hspace{1.65cm}  \Rightarrow \quad \mX_2 = \mX_1+1\,, \\
\label{eq:2HDMvevs}
& \mX_1 v_1^2 + \mX_2 v_2^2 =0  \quad  \Rightarrow \quad \mX_1 = - s^2_{\beta_1}\,. 
\end{align}
Since the quarks Yukawa operators are the same as in the previous model, 
the expression for the quark couplings in \eq{eq:C0} is the same,  
however now with $2 N=\mX_2-\mX_1=1$. It is 
now easy to see that, with $f_{ud}\approx 1/3$,  
the nucleophobic conditions \eqs{eq:CppCn}{eq:CpmCn} are satisfied at tree level in the parameter space region where  $\tan^2\beta_1 \approx 2$. 
Instead,  the electrophobic condition is not satisfied since the charge assignments give $C^0_e=\mX_1\neq0$. 
However, given that in this model the lepton 
charges are generation dependent, there are 
corrections to the mass eigenstate couplings due to lepton flavour mixing. Since in the lepton sector mixing effects can
be particularly large, as it was pointed out in Ref.~\cite{DiLuzio:2017ogq}  
electrophobia can still be enforced at the cost of a fine-tuned
cancellation yielding $C_e^0 + \delta_e^{\rm mix} \approx 0$.

\section{Astrophobic  axions beyond tree level}

\begin{figure}[t!]
\centering
\includegraphics[width=0.45\textwidth]{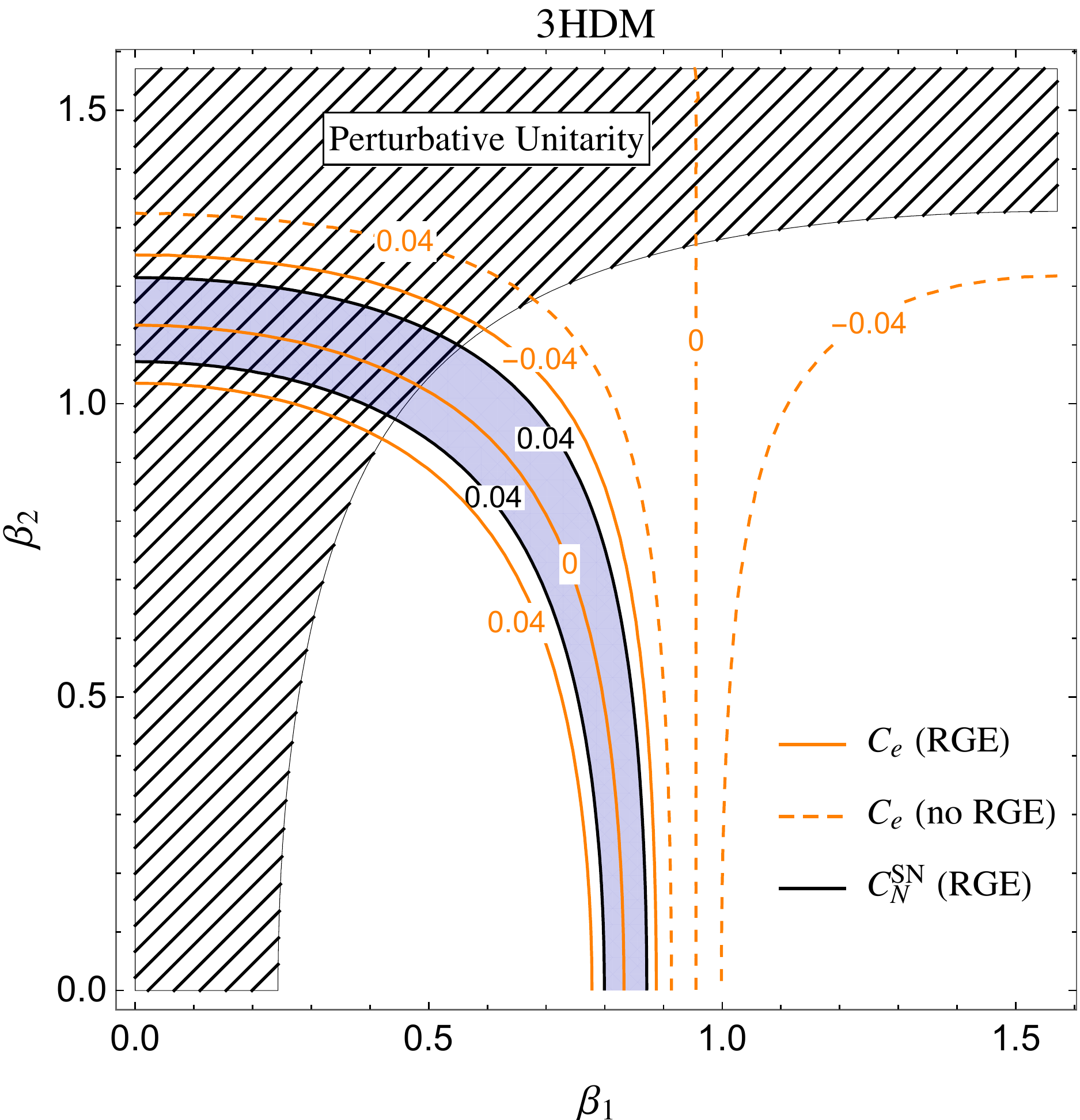}
\caption{
\label{fig:CN-Ce}
Contour lines for $C_e$ (orange) and $C^{\rm SN}_N$ 
(black, see text) in the 
$(\beta_1,\beta_2)$ plane for the astrophobic 3HDM. Solid lines 
include  RG corrections for $m_{\rm BSM}=10^{10}\GeV$,  dashed orange lines correspond to the tree-level results.}
\end{figure}
\begin{figure}[t!]
\centering
\includegraphics[width=0.45\textwidth]{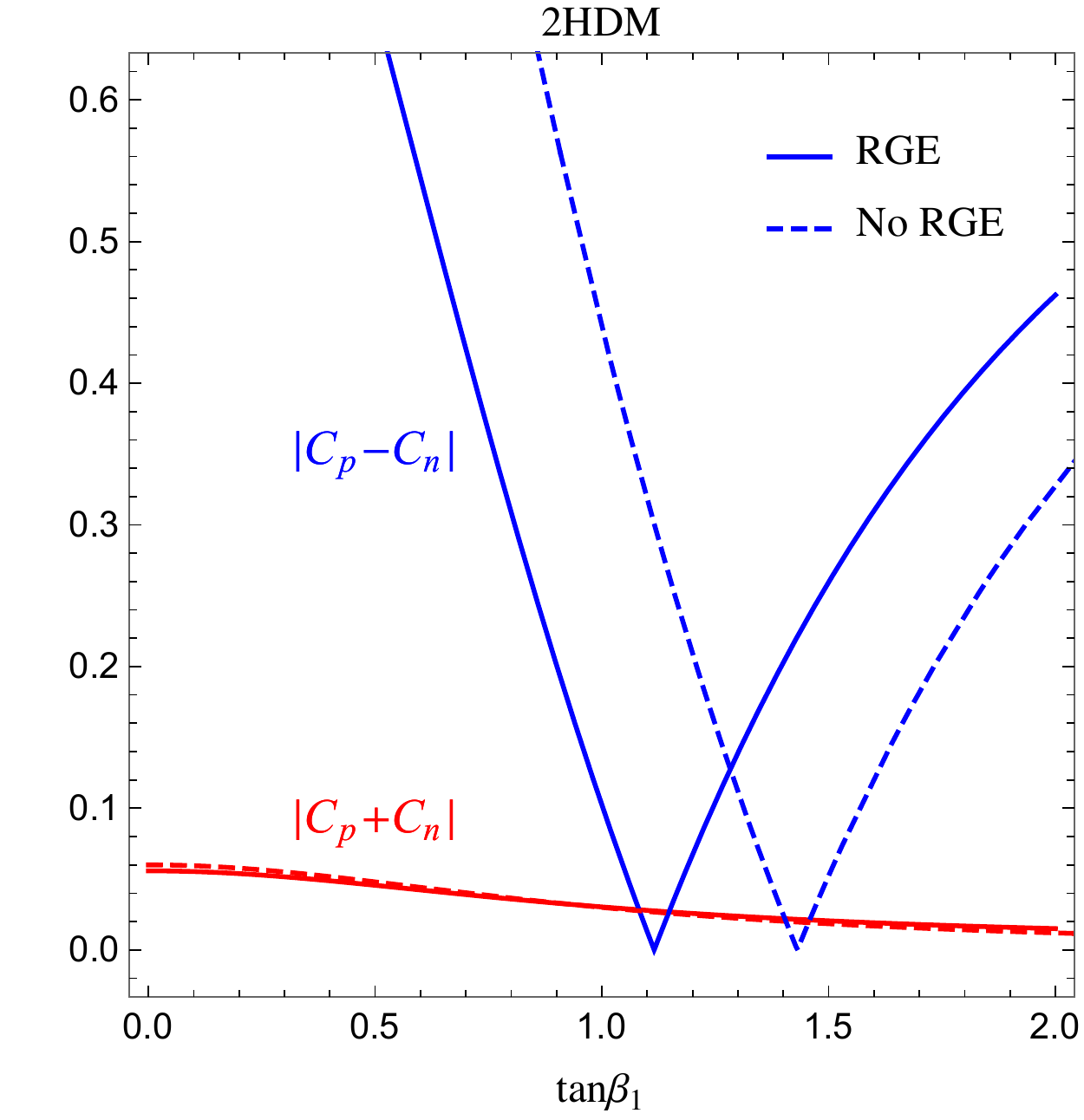}
\caption{The values of axion-nucleon couplings, 
$|C_p+C_n|$ (red) and $|C_p-C_n|$ (blue) in the nucleophobic 2HDM 
 as a function of $\tan\beta_1$. 
 Solid lines include  RG corrections
 for  $m_{\rm BSM} =10^{10}\GeV$, 
  dashed lines depict the tree-level results.
}
\label{fig:Cp-tanbeta}
\end{figure}

The leading  RG effects on the nucleo and electrophobic
conditions \eqs{eq:nucleo1}{eq:electro} can be 
understood from the formulae for the axion running couplings 
given in Eqs.~(\ref{eq:cazzoU1Y}).
The  top Yukawa coupling $Y_t$ gives the dominant contribution to the RH side 
of these equations. 
For the first generation fermions,
in the approximation in which all Yukawa couplings except $Y_t$ are neglected,
this contribution appears only through the last term 
 $\beta_\psi \,\gamma_H$ ($\psi= q_L,u_R,d_R,\ell_L,e_R$).   
 In this approximation the expression for $\gamma_H$ given in 
\eq{eq:Xdef} reduces to 
 $\gamma_H\approx 6Y^2_t (c'_{t_R} - c'_{t_L}) = 6 Y^2_t c^0_t$,
where $c^0_t$ denotes the axial-vector coupling of the top. 
We can now combine Eqs.~(\ref{eq:cazzoU1Y}) to  
obtain RG equations (RGEs) for the $u,d,e$ axial-vector couplings  $c_{u,d,e}$. Recalling the definition 
of the hypercharge ratio $\beta_\psi=Y_\psi/Y_H$,  it is easy to see that 
 the  $\gamma_H$  term will appear 
 in these equations 
 respectively  with coefficients 
$\beta_u-\beta_q = +1$ and $\beta_d-\beta_q=\beta_e-\beta_\ell=-1$.\footnote{The difference between the RH and LH hypercharge ratios  
is proportional to the weak-isospin of the LH component. This explains   
 the opposite sign between the $u$ and the $d,e$ coefficients.}
Hence, in this approximation we can write 
\begin{align}
\label{eq:Cu}
C_u & \approx C^0_u - \kappa_t \, C^0_t, \\ 
\label{eq:CdCe}
    C_{d,e}  & \approx C^0_{d,e} + \kappa_t\,  C^0_t\,,
\end{align}
where 
$C_{u,d,e} = c_{u,d,e}/(2N)$ 
are the couplings 
at the low scale $\mu$,  
$C^0_{u,d,e,t} = c^0_{u,d,e,t}/(2N)$
are the couplings at the high scale 
$f$
defined in terms of the PQ charges in \eq{eq:C0}, 
and the coefficient $\kappa_t \sim  6\, (Y_t/4\pi)^2 \log(m_{\BSM}/\mu)\,$ accounts for the running of the couplings  from the high 
scale $m_{\BSM}$ where the heavy Higgs components are integrated out,  
down to the low scale $\mu$. 

The first condition for nucleophobia 
is still satisfied by the running couplings due to the fact that the correction  proportional to  $\kappa_t$ cancels in the sum
\begin{align}
\label{eq:nucleo1RG}
 C_u+C_d \approx C_u^0+C_d^0 =1\,.   
\end{align}
RG effects modify instead the other two conditions \eqs{eq:nucleo2}{eq:electro}.
It is straightforward to see 
that now they are respectively satisfied 
for the following values of $\mX_3$: 
\begin{align}
    \mX_3 & = \frac{\frac{1}{2} (1-3 f_{ud})+\kappa_t}{1-\kappa_t}\, ,\\
        \mX_3 & = \frac{\kappa_t}{1-\kappa_t}\,.
\end{align}
We  see that the  same numerical accident that allows to 
enforce astrophobia with the tree-level relations in
\eq{eq:X3} (corresponding to $\kappa_t\to 0$) 
ensures that the same result still holds  after  including 
in the axion couplings the leading RG effects. 
Let us  note that this result is independent of the particular 
value of $\kappa_t$, that is, it does not depend on any specific value of the 
high scale  $m_\BSM$. Only the value of 
the PQ charges that realise the two conditions is affected by RG corrections, and while at tree level 
one has $\mX_3\approx 0$, for $\kappa_t\simeq 0.30$
one has instead $\mX_3 \approx 0.43$. 
Of course, since the PQ-hypercharge orthogonality condition in \eq{eq:PQ-Y}
is now satisfied for a non-vanishing value of $\mX_3$, 
the region in the $(\beta_1,\beta_2)$ plane where the axion 
can exhibit a remarkable degree of astrophobia
gets shifted accordingly, see Fig.\ref{fig:CN-Ce}.  
However, except for this modification 
in the viable  parameter space region, it is a remarkable result 
that  the astrophobic axion model introduced in Ref.~\cite{Bjorkeroth:2019jtx} still maintains its properties 
 after including RG corrections, without 
the need of any modification in the theoretical setup. 
Finally, it goes without saying that the 
nucleophobic property of the 2HDM  model in Ref.~\cite{DiLuzio:2017ogq}
are also preserved, but for a different VEVs ratio 
$\tan^2\beta_1 \approx 1.2$ (see  Fig.~\ref{fig:Cp-tanbeta}).   
Also the suppression of the axion-electron coupling 
can still be engineered, but with a corresponding shift in the value 
of the mixing correction $\delta_e^{\rm mix}$.

The results of this analysis, based on the approximate expressions \eqs{eq:Cu}{eq:CdCe},
are confirmed in Figs.~\ref{fig:CN-Ce} and \ref{fig:Cp-tanbeta}
that are obtained by numerically solving the full 
RGEs for the axion couplings given in  \ref{sec:AxionEFT}. 
In Fig.~\ref{fig:CN-Ce} we show 
the contour lines for different values 
of $C_e$ 
and $C^{\rm SN}_N=(C^2_n+0.61 C^2_p + 0.53 C_n C_p)^{1/2}$ 
in the $(\beta_1,\beta_2)$ plane. 
The latter combination of nucleon couplings
corresponds to the quadratic form which 
is bounded by the SN1987A neutrino burst duration~\cite{Carenza:2019pxu}. 
The  lowest value corresponds to 
$C^{\rm SN}_N\simeq0.02$ 
which is determined by the correction 
$\delta_s$ in \eq{eq:CppCn} 
(for comparison in the 
Kim-Shifman-Vainshtein-Zakharov (KSVZ) \cite{Kim:1979if,Shifman:1979if}
axion model 
$C^{\rm SN}_N = 0.36$). 
The hatched region in Fig.~\ref{fig:CN-Ce} 
denotes the perturbative unitarity bounds 
on the Yukawa couplings of the 3HDM 
(see e.g.~\cite{DiLuzio:2016sur,DiLuzio:2017chi}) 
translated in the $(\beta_1,\beta_2)$ plane. 
It is evident from Fig.~\ref{fig:CN-Ce} 
that, also in the case of running axion 
couplings,  electrophobia and nucleophobia occur in
overlapping regions, so that a single choice of the values
of the relevant parameters can simultaneously enforce 
all the astrophobic conditions.  Fig.~\ref{fig:Cp-tanbeta} instead 
displays the values of $C_p \pm C_n$  
as a function of $\tan\beta_1$ in the 2HDM case. 
As expected from the approximate expressions in 
\eqs{eq:Cu}{eq:CdCe}, running effects largely 
cancel out in the combination $C_p+C_n$, 
while they sizeably change the value of 
$\tan\beta_1$ for which  the  couplings combination 
$C_p-C_n$ is maximally suppressed from $\tan\beta_1 \simeq\sqrt{2}$ to $\tan\beta_1\simeq 1.1$.  
Nevertheless the same level of nucleophobia than in the tree level analysis 
can still be  obtained  regardless of the running 
effects.\footnote{We note in passing  that also the exponential 
enhancement of axion-nucleon couplings in the {\it nucleophilic} axion 
models of Ref.~\cite{Darme:2020gyx}  
is not spoiled by running effects. The reason being that the required 
cancellation between the QCD anomaly factors of first and second generation 
quarks holds at all orders.}

\begin{figure}[t!]
\centering
\includegraphics[width=0.45\textwidth]{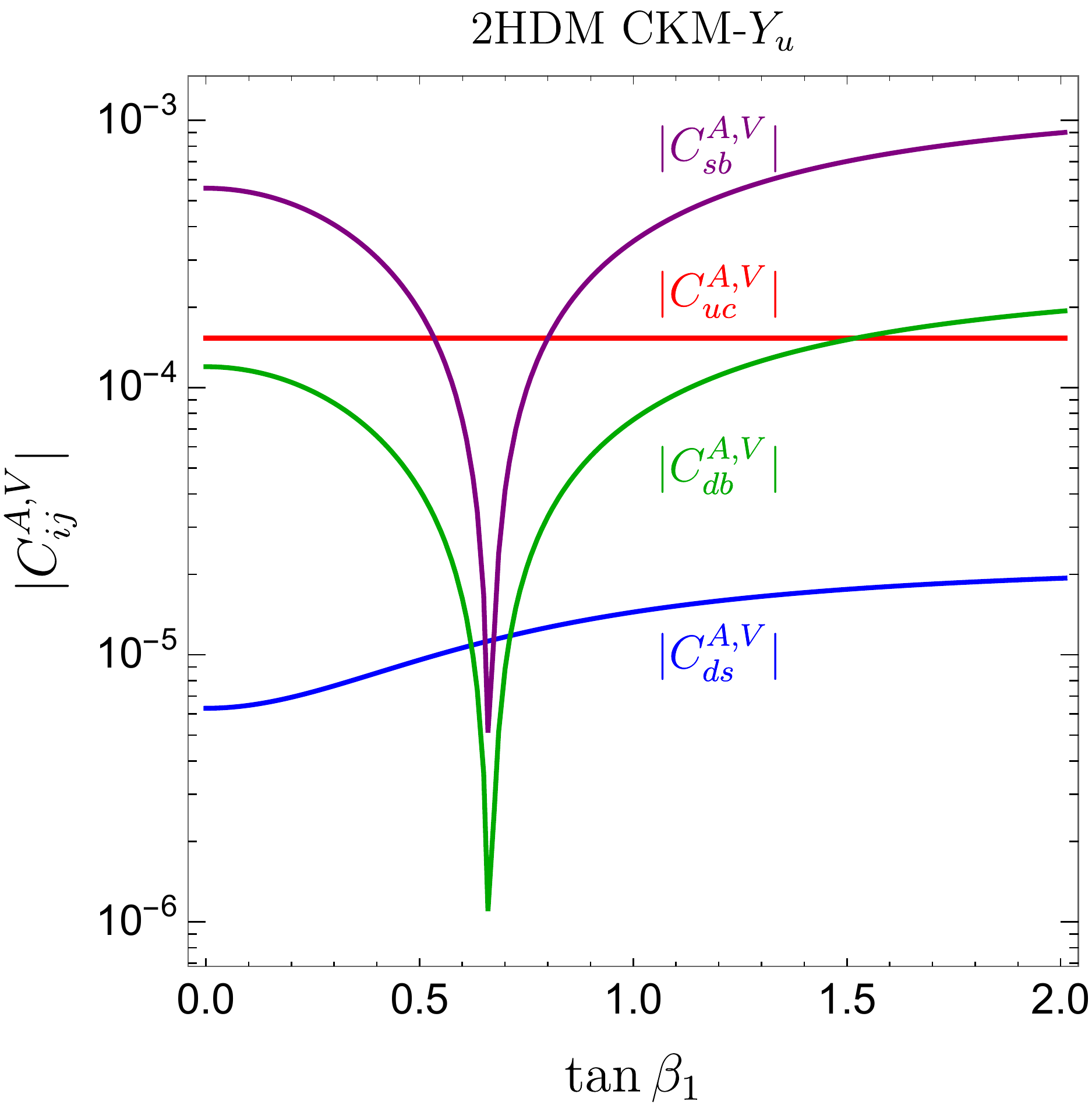} \ \ \ \
\caption{Flavour off-diagonal axion couplings 
$|(C^{A,V}_u)_{ij}|$ and $|(C^{A,V}_d)_{ij}|$ with $m_\BSM=10^{10}\GeV$ 
in the 2HDM for the CKM-$Y_u$. At tree level $(C^{A,V}_d)_{ij}=0$ but   
non-zero values arise radiatively, while  
$C^{A,V}_{uc} \neq 0$  but it does not receive  
RG corrections. 
\label{fig:Cfij}
}
\end{figure}

\section{Running effects on flavour-violating axion couplings}
\label{sec:offdiagonal1loop}

Flavour-violating axion couplings are generically expected in 
axion model with generation dependent PQ charge assignments,
and it is therefore important to study the 
 impact of RG corrections on these couplings. 
We focus for definiteness on the 
flavour off-diagonal 
 couplings between the axion and the quarks in the 
2HDM.
Since only the charges of the left-handed (LH) quarks are generation 
dependent (see \eq{eq:M1charges}) and recalling that 
$c_G = 2N = 1$,  using \eqs{eq:CfV}{eq:CfA} we can 
write\footnote{\eq{eq:CVCA} is defined at low energy, and thus  
it holds up to small corrections from right-handed (RH) 
mixings induced by running (see \eq{eq:cazzoU1Y}),
which lift the universality of the RH couplings. 
These effects are taken into account in the numerical analysis.}
\begin{align}
\label{eq:CVCA}
(C^V_{d/u})_{i\neq j} \approx - (C^A_{d/u})_{i\neq j} 
\approx (U_{d_L/u_L} c'_{q_L} U_{d_L/u_L}^\dag )_{ij} \, , 
\end{align}
where the LH rotation matrices $U_{d_L/u_L}$ are defined via
\beq
Y_u = U_{u_L}^\dagger \hat  Y_u  U_{u_R} \,, 
\quad
Y_d= U_{d_L}^\dagger \hat Y_d U_{d_R}  \,,
\label{eq:ckm}
\eeq
with $\hat Y_{u,d}$ the diagonal Yukawa matrices, and let us recall that   
 $U_{d_L/u_L}$ are related to the CKM matrix via  
$\vckm = U_{u_L} U_{d_L}^\dagger$.
Here we will consider the following two flavour ansatze: 
\begin{align}
\mbox{CKM-$Y_u$: } & Y_u = \vckm^\dag \hat Y_u \, , \quad Y_d = \hat Y_d \, , \quad (U_{u_L}=\vckm)  \, , \\
\mbox{CKM-$Y_d$: } & Y_u = \hat Y_u \, , \quad Y_d = \vckm \hat Y_d \, , \quad (U_{d_L}=\vckm^\dagger) \, . 
\end{align}
In the CKM-$Y_u$ case, $(C^{V,A}_{d})_{i\neq j}=0$ at the tree level and the non-zero $(C^{V,A}_{d})_{i\neq j}$ couplings are radiatively generated. 
We remark that the alignment of the flavour 
structure in the down sector 
is not  radiatively stable under the RG evolution,  
and hence  processes like  $K \to \pi a$  
can still occur with a rate sufficiently large to be 
observable.  
In the CKM-$Y_d$ case, $(C^{V,A}_{u})_{i\neq j}=0$ at the tree level, and it remains negligible, i.e.~at most  $O(10^{-9})$ even after including RG 
effects. 
For $f_a \gtrsim 10^{8}\,$GeV all the off-diagonal couplings remain well below the experimental limits reported in \Table{tab:limits2}, where the strongest constraint is $|C^V_{ds}|\leq 3.3\times 10^{-2} \times (f_a/10^{10}\,{\rm GeV})$ from Ref.~\cite{MartinCamalich:2020dfe}. 
\begin{table}[h!]
\centering
\begin{tabular}{|c|c|}
\hline
Coupling & Bound $[\, \times \ (f_a/10^{10}\GeV)]$\\ \hline
$|C_{uc}^V|$  & $\leq 2.1 \times 10^2$ \\ [2pt]
$|C_{ds}^V|$  & $\leq 3.3\times 10^{-2}$ \\ [2pt]
$|C_{db}^V|$  & $\leq 1.8 \times 10^2$ \\ [2pt]
$|C_{sb}^V|$  & $\leq 61$ \qquad \phantom{1}\\ [2pt]
\hline
$|C_{uc}^A|$  & $\leq 4.2 \times 10^2$ \\ [2pt]
$|C_{ds}^A|$  & $\leq 4.5 \times 10^2$ \\ [2pt]
$|C_{db}^A|$  & $\leq 1.5 \times 10^3$ \\ [2pt]
$|C_{sb}^A|$  & $\leq 8.7 \times 10^3$ \\ [2pt]
\hline
\end{tabular}
\caption{Current experimental bounds on axion flavour-violating couplings. See Ref.~\cite{MartinCamalich:2020dfe} for details.
\label{tab:limits2}
}
\end{table}

In the CKM-$Y_u$ case 
an interesting feature emerges (see  Fig.~\ref{fig:Cfij}). 
The $C_{qb}^{A,V}$ ($q=s,d$) couplings 
are strongly suppressed for $\tan\beta_1 \approx 0.65$.
This cancellation can be understood analytically
by keeping only leading top-loop effects. 
Employing the CKM-$Y_u$ structure and neglecting all Yukawa couplings except the top one, the RG evolution of the off-diagonal couplings can be cast in the form  
\beq
\label{eq:rgeCij}
\frac{d(c'_{q_L})_{i \neq j}}{d\log\mu} \propto \[ \frac{(c'_{q_L})_{ii}}{2}  + \frac{(c'_{q_L})_{jj}}{2}  - (c'_{t_R}) \] \,  Y_t^2 (\vckm^\dag)_{i3} 
(\vckm)_{3j} \, , 
\eeq
where only the diagonal couplings of $(c'_{q_L})_{ii}$ have been kept.   
Since both $(c'_{q_L})_{ii}$ and $(c'_{t_R})$ are positive, 
it is possible to cancel the quantity in the square brackets for $i=3$ or $j=3$ at a specific value $\tan\beta_1$. 
The RG corrections to $(c'_{q_L})_{i \neq j}$ are proportional to $(\vckm^\dag)_{i3} (\vckm)_{3j}$, which indicates that the off-diagonal axion couplings to the up-quarks do not receive the corrections, given that the CKM factors cancel out due to unitarity.  

In the CKM-$Y_d$ case,  on the other hand, 
flavour mixing occurs only through the down-quarks Yukawa couplings, and keeping only the 
top-loop contribution, the RG correction to the off-diagonal couplings  vanishes, 
namely $d(c'_{q_L})_{i \neq j}/d\log\mu \approx 0$. 
RG effects are thus captured solely by the running of the diagonal LH quark couplings $(c'_{q_L})_{ii}$ and matching corrections at the electroweak scale~\cite{Bauer:2020jbp}, which remain at the level of 1--\,4\,\%.

\section{Conclusions}
\label{sec:concl}

In this work we assessed the impact of RG effects on the axion couplings, 
focussing on the case of non-universal axion models. 
An important application of the RG analysis 
arises in the context  of the so-called astrophobic axions
of Refs.~\cite{DiLuzio:2017ogq,Bjorkeroth:2019jtx}, 
in which the axion couplings to nucleons and electrons 
can be simultaneously suppressed, 
thus allowing to relax the most stringent astrophysical constraints. 
In the original works the nucleo and electrophobic conditions 
were only set out at tree level, and it remained an important open question  
whether the conditions for astrophobia would still   
hold after including RG effects. In this paper we 
have shown that, perhaps unexpectedly,  the astrophobic features 
are not spoiled by  RG running of the axion couplings. 
The only effect  is a sizeable shift in the parameter space regions in which these conditions are realised. 

Since non-universal axion models necessarily imply  
certain flavour-violating axion couplings,   
we have also assessed the impact of running 
on these latter couplings. For instance, a tree level flavour  
structure aligned in such a way that off-diagonal couplings 
in the down sector are absent, is not stable under RG evolution, and  
we have estimated the irreducible  contributions  
to flavour violating processes arising from this type of effects. 

The tools developed in this work could be applied 
to other problems of phenomenological relevance. For instance, 
it could be interesting to see whether RG corrections 
can sizeably modify the fit to the so-called 
``stellar cooling anomalies'',  
improving on the 
tree-level analysis in Refs.~\cite{Giannotti:2017hny,DiLuzio:2021ysg}.

\section*{Acknowledgments}

The work of L.D.L. was partially supported by the European Union's Horizon 2020 research and innovation programme under the Marie Sk\l{}odowska-Curie grant agreement No 860881-HIDDEN.
E.N.~acknowledges support from 
a Mar\'ia de Maeztu  grant for a visit to the Institute of Cosmos Sciences,
Barcelona University, where this work was completed. 
E.N.~is supported in part by the INFN ``Iniziativa Specifica" Theoretical Astroparticle Physics (TAsP-LNF).
S.O.~and F.M.~acknowledges financial support from the State Agency for Research of the Spanish Ministry of Science and Innovation through the ``Unit of Excellence Mar\'ia de Maeztu 2020-2023" award to the Institute of Cosmos Sciences (CEX2019-000918-M), and from PID2019-105614GB-C21 and 2017-SGR-929 grants.

\appendix
\section{RGEs for axion EFTs}
\label{sec:AxionEFT}

In order to take into account running effects
it is convenient to adopt the Georgi-Kaplan-Randall (GKR) field basis \cite{Georgi:1986df}, 
where the PQ symmetry is realised non-linearly, so 
that under a $\U(1)_{\rm PQ}$ symmetry transformation 
all fields are invariant except the axion field, 
which changes by an additive constant 
$a \to a + \alpha f$, that is 
\begin{align} 
\label{eq:LaGKRH12}
&\mathcal{L}^{\rm GKR-2HDM}_a = \frac{1}{2} \partial_\mu a \partial^\mu a 
+ \sum_{A=G,W,B} c_A \frac{g_A^2}{32\pi^2} \frac{a}{f} F^A \tilde F^A \\
&\quad + \frac{\partial_\mu a}{f} 
\Big[ c_{H_1} H_1^\dag i \overleftrightarrow{D^\mu} H_1 + c_{H_2} H_2^\dag i \overleftrightarrow{D^\mu} H_2 
+ \bar q_L c_{q_L} \gamma^\mu q_L  \nonumber \\
&\quad + \bar u_R c_{u_R} \gamma^\mu u_R 
+ \bar d_R c_{d_R} \gamma^\mu d_R
+ \bar \ell_L c_{\ell_L} \gamma^\mu \ell_L 
+ \bar e_R c_{e_R} \gamma^\mu e_R 
\Big] \, , \nonumber
\end{align}
where $H_{1,2}^\dag \overleftrightarrow{D^\mu} H_{1,2} \equiv H_{1,2}^\dag (D^\mu H_{1,2}) - (D^\mu H_{1,2})^\dag H_{1,2}$
and $c_{q_L}, \ldots $ are diagonal matrices in generation space. 
Note that in the EFT we have neglected the heavy $\mathcal{O}(f)$ radial mode of $\Phi$   
and we focused for simplicity on the 2HDM  
(the generalization to an arbitrary number of Higgs doublets is 
straightforward). 
In order to match an explicit axion model 
to the effective Lagrangian in \eq{eq:LaGKRH12} at the high scale $\mu \sim \mathcal{O}(f)$, 
we perform an axion dependent field redenfinition: 
$\psi \to e^{-i \X_\psi a /f} \psi$, where $\psi$ spans over all the fields, and $\X_\psi$ is the corresponding PQ charge. 
Due to $\U(1)_{\rm PQ}$ symmetry, 
the non-derivative part of the renormalizable Lagrangian is 
 invariant upon this field redefinition, 
while the $d=5$ operators in 
\eq{eq:LaGKRH12} 
are generated 
from the variation of the kinetic terms and from the chiral anomaly. The couplings are then identified as  
\begin{align} 
\label{eq:cpsi}
c_{\psi} &= \X_\psi \, , \\
\label{eq:cA}
c_A & =
\sum_{\psi_R} 2 \X_{\psi_R} \Tr T^2_A(\psi_R) -
\sum_{\psi_L} 2 \X_{\psi_L} \Tr T^2_A(\psi_L)  
\, , 
\end{align}
where in the second equation 
$c_{\psi_{R,L}}$ refer to the charges of the chiral fermion fields.\footnote{Note that our anomaly coefficients $c_A$ have opposite sign with respect to those
in Refs.~\cite{Choi:2017gpf,Bauer:2020jbp,Choi:2021kuy}. This is due to the
fact that we are using a different convention for the Levi-Civita tensor, namely $\epsilon^{0123}=-1$.}
For the 2HDM   
introduced in \sect{sec:nonUnivAxion}, 
the charges $\X_\psi$,  
that can be read off from the Yukawa couplings in \eq{eq:2HDMleptons} can be set to
\begin{align}
\label{eq:M1charges}
\X_{q_i} &= (0, 0, \X_2 - \X_1)\, ,\
\X_{u_i} = - (\X_1, \X_1, \X_1)\, ,\: 
\X_{d_i} = (\X_2, \X_2, \X_2)\, , \nonumber \\
\X_{\ell_i} &=  -\X_{q_i} 
\, ,\:\X_{e_i}= - \mX_{u_i} 
\, ,
\end{align}
where $\mX_1 = -s^2_{\beta_1}$ and 
$\mX_2 = c^2_{\beta_1}$, 
see \eq{eq:2HDMvevs}, and we have  shifted the charges proportionally to $B$ and $L$ to set $\mX_{q_{1,2}} =\mX_{\ell_{1,2}}=0$. For the anomaly coefficients in  
 \eq{eq:cA} one has $(c_G, c_W, c_B) = (1, -2, 8/3)$
and, in particular, the electromagnetic to QCD 
anomaly ratio is $E/N \equiv (c_W + c_B) / c_G = 2/3$. 
For the 3HDM 
instead the lepton charges are $\mX_\ell=0,\, \mX_e=\mX_3$, the 
corresponding anomaly coefficients read 
 $(c_G, c_W, c_B) = (3,-9,17)$  and $E/N=8/3$.

Running effects induced by Yukawa couplings (and in particular by the Yukawa of the top 
which are the most relevant ones) 
only occur below the scale of the heavy radial modes of the 2HDM, 
that will be denoted as 
$m_{\rm BSM} \simeq m_{H,\,A,\,H^\pm}$,  with the heavy scalars  assumed to be degenerate in the decoupling limit (see e.g.~\cite{Gunion:2002zf}). 
This is due to the fact that as long as the complete set  of Higgs doublets  appear in the EFT, 
the PQ current is conserved (up to anomalous effects) and thus the couplings, which correspond to PQ charges, do not renormalize. Once the heavy scalar components are 
integrated out,  the sum rule of PQ charges set by $\U(1)_{\rm PQ}$ invariance  breaks down,  and non-vanishing contributions to 
the running of the couplings arise (see e.g.~\cite{Choi:2021kuy}). 
We can now directly match \eq{eq:LaGKRH12} 
at the scale $\mu = \mathcal{O}(m_{\rm BSM})$
with a GKR  basis  featuring  only one SM-like Higgs doublet 
\begin{align} 
\label{eq:LaGKRH}
\mathcal{L}^{\rm GKR-SM}_a &= \frac{1}{2} \partial_\mu a \partial^\mu a 
+ \sum_{A=G,W,B} c_A \frac{g_A^2}{32\pi^2} \frac{a}{f} F^A \tilde F^A \\
&+ \frac{\partial_\mu a}{f} 
\Big[ c_{H} H^\dag i \overleftrightarrow{D^\mu} H  
+ \bar q_L c_{q_L} \gamma^\mu q_L  \nonumber \\
&+ \bar u_R c_{u_R} \gamma^\mu u_R 
+ \bar d_R c_{d_R} \gamma^\mu d_R
+ \bar \ell_L c_{\ell_L} \gamma^\mu \ell_L 
+ \bar e_R c_{e_R} \gamma^\mu e_R 
\Big] \, , \nonumber
\end{align}
where $c_{H} = c_{H_1} c^2_\beta + c_{H_2} s^2_\beta$,  
which follows from the projections 
on the SM Higgs doublet: 
$H_1 \to c_\beta \, H$ and  
$H_2 \to s_\beta \, H$, 
consistently with the definition 
of $\tan\beta\equiv \tan\beta_1 = v_2/v_1$. 
In particular, by employing global $\U(1)_Y$ invariance, 
it is convenient to cast 
the RGEs in a 
form that does not depend explicitly on $c_H$. 
This can be achieved via  
the axion-dependent field redefinition:  
$\psi \to \psi' = e^{-i c_H \beta_\psi a/f} \psi$,  
with $\beta_\psi = Y_\psi / Y_H$
the ratio of the corresponding hypercharges, 
which redefines the effective couplings as $c'_\psi = c_\psi - c_H \beta_\psi$ 
(so in particular $c'_H = 0$). 
In this basis the RGEs read: 
 \begin{align}
\label{eq:cazzoU1Y}
(4\pi)^2 \frac{dc'_\qL}{d\log\mu} 
           & = \frac{1}{2} \{ c'_\qL, Y_u Y_u^\dagger + Y_d Y_d^\dagger \} 
                   - Y_u c'_\uR Y_u^\dagger - Y_d c'_\dR Y_d^\dagger \nonumber\\
           &+ \left( 8 \alpha_s^2 \widetilde{c}_G + \frac{9}{2} \alpha_2^2 \widetilde{c}_W 
                  + \frac{1}{6} \alpha_1^2 \widetilde{c}_B \right) \, \1  -\b_q \, 
\gamma_H \, \1
\, , \nonumber\\
(4\pi)^2 \frac{dc'_\uR}{d\log\mu} 
           & = \{ c'_\uR, Y_u^\dagger Y_u \} - 2 Y_u^\dagger c'_\qL Y_u 
                  - \left( 8 \alpha_s^2 \widetilde{c}_G + \frac{8}{3} \alpha_1^2 \widetilde{c}_B \right) \, \1 \nonumber \\ 
                  &-\b_u \, 
\gamma_H \, \1
                  \, , \nonumber\\
(4\pi)^2 \frac{dc'_\dR}{d\log\mu} 
           & = \{ c'_\dR, Y_d^\dagger Y_d \} - 2 Y_d^\dagger c'_\qL Y_d 
                  - \left( 8 \alpha_s^2 \widetilde{c}_G + \frac{2}{3} \alpha_1^2 \widetilde{c}_B \right) \, \1 \nonumber \\ 
                  &-\b_d \, 
\gamma_H \, \1
\, , \nonumber\\
(4\pi)^2 \frac{dc'_\lL}{d\log\mu} 
          & = \frac{1}{2} \{ c'_\lL, Y_e Y_e^\dagger \} - Y_e c'_\eR Y_e^\dagger  
                  + \left( \frac{9}{2} \alpha_2^2 \widetilde{c}_W + \frac{3}{2} \alpha_1^2 \widetilde{c}_B \right) \, \1 \nonumber \\ 
                  &-\b_\ell \, 
\gamma_H \, \1
\, , \nonumber \\
(4\pi)^2 \frac{dc'_\eR}{d\log\mu} 
          & = \{ c'_\eR, Y_e^\dagger Y_e \} - 2 Y_e^\dagger c'_\lL Y_e  
                  - 6 \alpha_1^2 \widetilde{c}_B \, \1-\b_e \, 
\gamma_H \, \1
\, , 
\end{align}
where 
\begin{align}
\label{eq:Xdef}
\gamma_H
           & =
                   - 2 \, \tr\( 3Y_u^\dagger c'_\qL Y_u - 3Y_d^\dagger c'_\qL Y_d - Y_e^\dagger c'_\lL Y_e \) \nonumber \\
&+ 2 \, \tr\( 3Y_u c'_\uR Y_u^\dagger - 3Y_d c'_\dR Y_d^\dagger - Y_e c'_\eR Y_e^\dagger \) \, , \nonumber \\
\widetilde{c}_G
           & = c_G - \tr\( c'_{u_R} + c'_{d_R}- 2c'_{q_L}  \) \, , \nonumber\\
\widetilde{c}_W
           & = c_W + \tr\( 3c'_{q_L} + c'_{\ell_L} \) \, , \nonumber\\
\widetilde{c}_B
           & = c_B - \tr\( \frac{1}{3} ( 8c'_{u_R} + 2c'_{d_R} -c'_{q_L} ) + 
           2 c'_{e_R}- c'_{\ell_L} \) \, .
\end{align}
Note that the $c_A$ ($A = G,W,B$) Wilson coefficients in \eq{eq:Xdef} do not run,   
since in the normalization of \eq{eq:LaGKRH12} the scale dependence of the operator $a F^A\tilde F^A$  
is  accounted for by the running of the gauge 
couplings \cite{Bauer:2020jbp,Chetyrkin:1998mw}.  

\eq{eq:LaGKRH} is matched at the scale $\mu = \mathcal{O}(m_Z)$ with the 
$\SU(3)_C \times \U(1)_{\rm EM}$-invariant
axion effective Lagrangian below the electroweak 
scale
\begin{align}
\label{eq:LeffEW}
\mathcal{L}_a &\supset 
\frac{g_s^2}{32\pi^2} \frac{a}{f_a} G \tilde G + 
\frac{c_\gamma}{c_G} \frac{e^2}{32\pi^2} \frac{a}{f_a} F \tilde F \nonumber \\
&+ \sum_{f = u,\,d,\,e} \frac{\partial_\mu a}{2 f_a} \bar f_i \gamma^\mu \left( (C^V_f)_{ij} + (C^A_f)_{ij} \gamma_5 \right) f_j \, , 
\end{align}
where we have introduced the standard QCD  normalization 
factor for the $aG\tilde G$ term and defined 
the axion decay constant $f_a = f / c_G$, while 
$c_\gamma = c_W + c_B$. We further have  
\begin{align}
\label{eq:CfV}
C^V_f &= \frac{1}{c_G} ( U_{f_R} c'_{f_R} U_{f_R}^\dag + U_{f_L} c'_{f_L} U_{f_L}^\dag ) \, , \\
\label{eq:CfA}
C^A_f &= \frac{1}{c_G} ( U_{f_R} c'_{f_R} U_{f_R}^\dag - U_{f_L} c'_{f_L} U_{f_L}^\dag ) \, , 
\end{align}
where $U_{f_{L,R}}$ are  the unitary matrices that diagonalize the SM fermion mass matrices, and 
$c'_{u_L} = c'_{d_L}  = c'_{q_L}$.  
After including matching corrections 
at the weak scale \cite{Bauer:2020jbp}, 
the running for $\mu < m_Z$ is given by 
\begin{align}
(4\pi)^2 \frac{d(C^A_{u})_{ii}}{d\log\mu} 
           & = - 16 \alpha_s^2 \wt{c}_G - \frac{8}{3} \alpha_{\rm em}^2 \wt{c}_\gamma \, , \nonumber\\
(4\pi)^2 \frac{d(C^A_{d})_{ii}}{d\log\mu} 
           & = - 16 \alpha_s^2 \wt{c}_G -  \frac{2}{3} \alpha_{\rm em}^2 \wt{c}_\gamma \, , \label{eq:RGEs} \nonumber \\
(4\pi)^2 \frac{d(C^A_{e})_{ii}}{d\log\mu} 
           & = - 6 \alpha_{\rm em}^2 \wt{c}_\gamma \, , 
\end{align}
with 
\begin{align}
\wt{c}_G(\mu) &= 1 - \sum_q C^A_q(\mu) \Theta(\mu-m_q) \, , \\
\wt{c}_\gamma(\mu) &= \frac{c_\gamma}{c_G} - 2 \sum_f N_c^f Q_f^2 C^A_f(\mu) \Theta(\mu-m_f) \, ,
\end{align}
where $\Theta(x)$ is the Heaviside theta function, while 
$N_c^f$ and $Q_f$ denote respectively the colour number and EM charge of the fermion $f$. 
Note that the off-diagonal couplings $(C^{A,V}_{f})_{i\neq j}$ do not run below the electroweak scale, 
while the diagonal vector couplings $(C^{V}_{f})_{ii}$ can be 
set to zero thanks to the conservation of the vector current.

The axion-nucleon couplings, neglecting the tiny contributions of the matrix elements 
 $\Delta_{t,b,c}$ of the heavy flavours, can be calculated by using 
\begin{align}
\label{eq:CpDelta}
C_p &=
C_u \Delta_u + C_d \Delta_d + C_s \Delta_s 
-\left(\frac{m_d \Delta_u}{m_u+m_d} + \frac{m_u \Delta_d}{m_u+m_d} \right) \, ,
\\
\label{eq:CnDelta}
C_n &=
C_d \Delta_u + C_u \Delta_d + C_s \Delta_s 
-\left(\frac{m_u \Delta_u}{m_u+m_d} + \frac{m_d \Delta_d}{m_u+m_d} \right) \, ,
\end{align}
where $C_{u,d,s} = C^A_{u,d,s}(2\GeV)$ 
(we neglect here for simplicity model-dependent tree-level flavour mixing effects -- see \eq{eq:CfA})
are evaluated by numerically solving the RGEs, Eqs.~(\ref{eq:cazzoU1Y}) and (\ref{eq:RGEs}), 
starting from the boundary conditions set at the scale $f$ 
(cf.~below \eq{eq:cA}). 
In \eqs{eq:CpDelta}{eq:CnDelta}
$\Delta_{u,d,s}$ represent the nucleon matrix elements of 
the light quarks axial-vector current, whose numerical values are  
 $\Delta_u = 0.897(27)$, $\Delta_d = -0.376(27)$, 
$\Delta_s =  -0.026(4)$, while $m_u(2 \GeV)/m_d(2 \GeV)=0.48(3)$ \cite{diCortona:2015ldu}.
With these inputs, we arrive at 
\begin{align}
\label{eq:CpAppB}
C_p &= 0.90 C_u - 0.38 C_d - 0.03 C_s - 0.48
\, , \\
\label{eq:CnAppB}
C_n &= 0.90 C_d - 0.38 C_u - 0.03 C_s - 0.04
\, .
\end{align}
In the calculation, we have employed the two-loop running 
for gauge and Yukawa couplings, and the input values for the SM Yukawa and CKM mixings are extracted from Ref.~\cite{Antusch:2013jca}.

\bibliographystyle{elsarticle-num} 
\bibliography{bibliography}

\end{document}